\documentclass[english]{article}
\usepackage[T1]{fontenc}
\usepackage[latin9]{inputenc}
\usepackage{geometry}
\usepackage{color}
\usepackage{amssymb}
\usepackage{graphicx}

\makeatletter
\newcommand{\lyxaddress}[1]{
\par {\raggedright #1
\vspace{1.4em}
\noindent\par}
}

\makeatother

\usepackage{babel}
\begin{document}

\title{Experimental quantum mechanics in the class room: Testing basic ideas
of quantum mechanics and quantum computing using IBM quantum computer}

\author{Anirban Pathak}
\maketitle

\lyxaddress{\begin{center}
Jaypee Institute of Information Technology, A10, Sector 62, Noida,
UP 201307
\par\end{center}}
\begin{abstract}
Since the introduction of quantum mechanics, it has been taught mostly
as a theoretical subject. It is also viewed as a theory that provides a
best understanding of the nature, but which does not have much practical
applications in our day to day life. This notion has been considerably
changed in the recent past with the advent of quantum information
processing. Further, recently, IBM has introduced a set of quantum
computers that are placed in the cloud and can be accessed freely
from your class room though your mobile phones, PCs, Laptops having
an internet connection. In this article, we will show that the IBM
quantum computer can be used to demonstrate many fundamental concepts
of quantum mechanics, quantum computing and communication. 
\end{abstract}

\section{Introduction}

The journey of quantum physics started with the introduction of Planck's
law in 1900. For about a quarter century, many seminal works, e.g.,
Einstein's work on the photoelectric effect (1905), Bohr model (1913),
Compton effect (1923), de Broglie's work on matter wave (1924), etc.,
enriched quantum physics and led to the foundation of quantum mechanics.
Finally, in 1925, quantum mechanics was introduced. Specifically,
in mid 1925, Heisenberg introduced matrix mechanics and in late 1925,
Schrodinger introduced wave mechanics, which was complemented soon
by the extremely convincing experiment of Davisson and Germer (1927)
and the beautiful idea of Heisenberg (1927), which is now known as
Heisenberg's uncertainty principle. Many of the ideas associated with
quantum mechanics was counter intuitive and without having any classical
analogue. Naturally, many scientists (including Einstein) often tried
to criticize it, and others (including Bohr) tried to defend it. This
healthy debate led to a probabilistic interpretation of quantum mechanics
(now known as Copenhagen interpretation). In the text books, we usually
follow this interpretation of quantum mechanics. However, all the
issues with the interpretation of quantum mechanics are not yet settled
and still scientists work on them under the domain of quantum foundation.
In this article, we are not going to discuss foundational issues associated
with the quantum mechanics. Rather, we would like to stress on the
fact that there are features of quantum mechanics which distinguishes
it from the classical physics (some of them are mentioned in Section
\ref{subsec:Some-basic-features}), and we don't observe the direct manifestation
of these features in our day to day life. Consequently, they appear
to be counter intuitive to us, and often students find it difficult
to accept\footnote{This is not surprising as even some of the founder fathers of the
subject found it difficult to accept.}. Usually, in the undergraduate courses quantum mechanics is taught
as a theoretical subject without any laboratory component. So students,
are somehow compelled to accept the features of quantum mechanics
as it is told by the teachers or written in the text books. In contrast
to this scenario, a recent initiative by IBM (which is of course not
designed for the class room teaching) has opened up the possibility
of doing some simple experiments in the class room using IBM quantum
computers that are placed in cloud by IBM corporation and can be accessed
freely by the students through their mobile phones, laptops, etc.,
provided there is an access to the internet in the class room. In this
article, we will elaborate on how to perform simple experiments using
IBM quantum computer in the class room to illustrate the features
of quantum mechanics that distinguishes it from the classical theories.
Specially, we will concentrate on the features like collapse on measurement, the
probabilistic nature of quantum mechanics and the existence of entanglement.
We will also state some applications of quantum mechanics that are
directly connected with the performed experiments. 

As we mentioned, ideas of quantum mechanics will be illustrated using
quantum computers. Naturally, you can guess that these are the computers
which work using the principle of quantum mechanics. The notion of such
computers was introduced in 1980s, and that led to the birth of a
subject called quantum computing. Almost around the same time quantum
communication also started its journey. The beauty of quantum computing
lies in its computing power- a quantum computer can perform various
tasks much faster than any of its classical counterparts. Similarly,
in quantum communication we can perform certain communication tasks
that are not doable using classical resources. For example, quantum
cryptography provides unconditional security- an extremely desirable
feature for secure communication, but not achievable in the classical
world. 

Although, a quantum computer is proven to be more powerful than its
classical counterparts, it's not easy to build as quantum states interact
with its environment and often collapse or get modified. It's specially
difficult to build large quantum computers, but even making a small
quantum computer is so difficult that most universities and colleges
don't have any quantum computer. Now, IBM's recent initiative- IBM
quantum experience \cite{IBM} is providing access of small quantum
computers to everyone as the quantum computers are placed in cloud
and its access is free. IBM quantum experience would provide the backbone
of this article. 

To make this article an almost self sufficient reading material, the
rest of the article is organized as follows. In Section \ref{sec:Visualization-of-the},
we briefly introduce the readers with the conceptual visualization
of the computing devices and establish that every gate is a small
computing device and the standard optical elements can be visualized
as gates (i.e., as small computing devices). As optics helps us greatly
in visualizing quantum mechanics and it is also taught in undergraduate
courses, we elaborate on some ideas of quantum mechanics through optics
in Section \ref{subsec:Understanding-the-concepts}, but prior to
that in Section \ref{subsec:Some-basic-features}, we describe basic
features of quantum mechanics and how are those connected to the basic
building blocks of quantum computing. In Section \ref{sec:IBM-Quantum-Computer},
we describe the technical aspects of IBM quantum computer and how
to use it. In Section \ref{sec:Performing-simple-experiments}, we
describe how to perform simple experiments (that verifies fundamental
characteristics of quantum mechanics) in the class room using IBM
quantum experience. The beginners and matured readers can skip the previous sections and can play with experiments after reading Section \ref{sec:Performing-simple-experiments}. Finally, the article is concluded in Section \ref{sec:Conclusion}.

\section{Visualization of the working of a small computing device\label{sec:Visualization-of-the}}

Let's start our discussion with a simple question: What is a computer?
It is (usually, but not always) an electronic device that can perform
computation using a set of received information (data) and produce
a result which can be considered as the output. In brief, it uses
input states, manipulates them by following certain rules and thus
performs computation and finally yields the output of the performed
computation. Now this answer leads to a question: which component
of the computer actually performs the computation? Is it the key-board,
monitor, battery or something else? The computing task is performed
by the ICs, which are very large circuits. These large circuits are
made of smaller circuits, and each of the smaller circuits is made
of some gates. We are familiar with the conventional irreversible
gates, e.g., NAND, NOR, OR, AND. A closer look into them would reveal
that each of these gates actually computes a function. For example,
NAND gate which is a universal gate computes a function $f(x,y)=z:\,\,z=\overline{xy}\,{\rm and}\:x,y,z\in\left\{ 0,1\right\} .$
Similarly, a NOR gate computes a function $f(x,y)=z:\,\,z=\overline{x+y}\,{\rm and}\:x,y,z\in\left\{ 0,1\right\} $
and an AND gate computes $f(x,y)=z:\,\,z=xy\,{\rm and}\:x,y,z\in\left\{ 0,1\right\} .$
Thus, each gate computes a function, and in principle they can be
viewed as the tiny computers or just as a small computing device.
Now, if we combine, a few of such gates, we would obtain a small circuit
which would also perform a computing task, usually it would be able
to perform a computing task more involved than the computing tasks
performed by the individual gates. Consequently, we can visualize
circuits as small computing devices containing a few gates. An appropriate
combination of many such small computing devices (circuits) would
lead to the design of ICs and the combination of couple of ICs would
lead to the architecture of a modern computer. In this article, in
what follows, we will consider individual gates and circuits (both
classical and quantum) as computing devices for the obvious reason
that each of them computes a function. Further, in what follows, we
will show that common optical elements like beam splitter (BS) and mirror
can be viewed as quantum gates and their well known combinations (e.g.,
Mach-Zehnder interferometer (MZI) with a source having extremely low intensity)
can be viewed as quantum circuits or a very small quantum computer. 

\section{Let's recall a bit of quantum mechanics and optics \label{sec:Let's-recall-a}}

In this section, we will briefly recall some elementary ideas of quantum
mechanics and optics that are usually taught in the undergraduate
courses. It's expected that the readers are familiar with these concepts.
However, they are briefly mentioned here to provide a kind of completeness
to this article. IBM quantum computers are not based on optics. However,
we have briefly discussed the optics as that would help us to map
the simple experiments to be performed with the IBM quantum computers
with the concepts of quantum mechanics taught in the undergraduate
classes.

\subsection{Some basic features of quantum mechanics \label{subsec:Some-basic-features}}

It is not our purpose to discuss the axiomatic structure of quantum mechanics
or to describe the postulates of quantum mechanics. We would prefer
to suggest the interested readers to follow any standard text book
for them. Here, we will just mention a set of important features of
quantum mechanics that distinguish it from a classical theory and
would help us in understanding the content of the rest of the article.
\begin{itemize}
\item Anything that you can measure is called a physical observable in the
classical theory. For every physical observable there is an operator
in the quantum mechanics. \\
Comment: You can see me, but I am not an observable as you cannot
measure me. However, you can measure my age, momentum, position, etc.,
so time, momentum and position are physical observables in a classical
theory and there exist unique operators for them in the quantum mechanics.
For example, momentum in $X$- direction is represented by the operator
$\hat{p}_{x}=-i\hbar\frac{\partial}{\partial x},$ similarly the operator
for energy is $\hat{E}=i\hbar\frac{\partial}{\partial t}$. Lucidly
speaking, in quantum mechanics you have operators corresponding to
measurable properties.
\item Any such operator $\hat{A}_{op}$ would satisfy eigen value equation
of the form $\hat{A}_{op}\psi=\lambda\psi,$ while operate on the
wave function $\psi.$ Eigen values $(\lambda)$ are discrete, and
these are the values that can be obtained on measuring the property
associated with the operator $\hat{A}_{op}.$ The word quantum actually
means discrete. Now a measurement cannot yield an imaginary number.
Just think that you have never heard of (3+4i) kg sugar or (16-9i)
mm long noodles. This demand of real outcome of the measurement implies
hermiticity of the quantum operators that correspond to the physical
properties. In other words, it ensures that all such operators would
satisfy $\hat{A}_{op}^{\dagger}=\hat{A}_{op}.$ 
\item The wave function $\psi$ is obtained as the solution of a particular
eigen value equation. To be precise it is the eigen function of energy
operator and the corresponding equation is referred to as Schrodinger
equation which is usually described in two forms- time independent
Schrodinger equation (for convenience we are writing 1 dimensional
equation) $\left[-\frac{\hbar^{2}}{2m}\frac{d^{2}}{dx^{2}}+V(x)\right]\psi(x)=E\psi(x),$
and time dependent Schrodinger equation $i\hbar\frac{\partial\psi}{\partial t}=\hat{H}\psi$,
where $\hat{H}$ describes the Hamiltonian of the system. From the
time dependent Schrodinger equation, we can easily obtain the relation
$\psi(t)=\exp\left(-\frac{i\hat{H}}{\hbar}t\right)\psi(0),$ which
describes the time evolution of the wave function. Now, as the operator
$\hat{H}$ corresponds to a physical observable, it must be Hermitian
and satisfy $\hat{H}=\hat{H}^{\dagger}.$ Consequently, the operator
$\hat{U}=\exp\left(-\frac{i\hat{H}}{\hbar}t\right)$, which describes
the time evolution of the wave function would satisfy the condition
of unitarity as $\hat{U}^{\dagger}=\exp\left(\frac{i\hat{H}^{\dagger}}{\hbar}t\right)=\exp\left(\frac{i\hat{H}}{\hbar}t\right)=\hat{U}^{-1}$.
Unitary operators are not always Hermitian. They are Hermitian if
and only if they are self-inverse. The proof is obvious, as for a
self-inverse operator we have $\hat{U}=\hat{U}^{-1}$ and for a unitary
operator we have $\hat{U}^{-1}=\hat{U}^{\dagger}$. So, a self-inverse
unitary operator would satisfy $\hat{U}=\hat{U}^{\dagger}$, the condition
of Hermiticity. \\
Now consider the quantum state $\psi(0)$ as the initial (input) state
and $\psi(t)$ as the final (output) state. This consideration would
show that we can visualize the operator $\hat{U}=\exp\left(-\frac{i\hat{H}}{\hbar}t\right)$
as a quantum gate which maps an input state into an output state by
following a certain transformation rule. This is consistent with the
conventional definition of gates, and thus, we can conclude that a
suitably chosen Hamiltonian and the evolution time can be used to
build a desired quantum gate. 
\item If we look at the time independent Schrodinger equation, we can easily
realize that it is a linear differential equation. We know that a
superposition of solutions of linear differential equation is also
a solution of the equation. Therefore, a linear combination of all
the valid solutions (wave functions) will also be a valid solution
of the Schrodinger equation. Up to this point, there is nothing quantum
mechanical. Beauty and mystery of quantum mechanics arise with the
introduction of a counter intuitive property that tells you that a
quantum state remains in the above type of superposition state until
it is measured, but as soon as you perform a measurement it collapses
to one of the possible states. For example, consider that you have
a 2 level atom which is in a state $\psi=\alpha\psi_{ground}+\beta\psi_{excited}.$
If you perform a measurement, the state will collapse to $\psi_{ground}$
with probability $|\alpha|^{2}$ and to $\psi_{excited}$ with probability
$|\beta|^{2}$, where $|\alpha|^{2}+|\beta|^{2}=1.$ Here, it may
be noted that on measurement quantum state $\psi$ collapses to one
of the possible states completely randomly- no one, not even a super hero
can predict whether $\psi$ will collapse to $\psi_{ground}$ or
to $\psi_{excited}$. This property can be used to make a true random
number generator. In fact, commercial quantum random number generators
exist- a famous one is known as QUANTIS which is based on this principle
and is marketed by IdQuantique. A brief description of this will be
given later.
\end{itemize}

\subsubsection{Basic building blocks of a quantum computer: qubits, quantum gates
and quantum circuits}

In the aforementioned quantum state $\psi=\alpha\psi_{ground}+\beta\psi_{excited}$,
we may refer to the ground state as ``0'' and the excited state
as ``1''. Then the beauty of the quantum state $\psi=\alpha\psi_{ground}+\beta\psi_{excited}$
(a feature that distinguishes it from a classical state) would be
the fact that the quantum state can be simultaneously in state ``0''
and ``1'', but on measurement would collapse to one of these states.
Such a quantum state of a two level system is called a ``qubit''
or quantum bit in analogy with the familiar notion of bit. In the
classical world, any two level physical system can be used to represent
a bit, where one level would represent ``0'' and the other level
would represent ``1'', so the system will be either in ``0'' or
in ``1''. In contrast, in the quantum world the system can be in
the superposition state. Many of the advantages of quantum world that
establishes quantum supremacy actually arises from this quantum superposition
phenomenon. By now, it must be clear that the aforementioned two-level
atom is just an example of a qubit. A qubit can be realized in various
ways- we just need a two-level quantum system. It can be a photon
passing through a BS (let us refer the reflected path as
"1'' and the transmitted path as "0''), a nucleon having arbitrary
spin (spin up being "0'' and down being "1''), and so on. 

For the sake of convenience, in what follows, we may use a notation
known as Dirac's bra-ket notation. In this notation, $\psi_{ground}$
(or a state in a two level quantum system that represents ``0'')
is written as $|0\rangle=\left[\begin{array}{c}
1\\
0
\end{array}\right]$ (read $|0\rangle$ as ket-zero). Similarly, the other state that
represents ``1'' is written as $\psi_{excited}=|1\rangle=\left[\begin{array}{c}
0\\
1
\end{array}\right]$ (read $|1\rangle$ as ket-one). Now we can write the qubit as $|\psi\rangle=\alpha\left[\begin{array}{c}
1\\
0
\end{array}\right]+\beta\left[\begin{array}{c}
0\\
1
\end{array}\right]=\left[\begin{array}{c}
\alpha\\
\beta
\end{array}\right].$ In this notation, transpose conjugate of $|\psi\rangle$ is written
as $\langle\psi|=[\alpha^{*}\,\,\,\beta^{*}]$ and pronounced as bra-$\psi.$
Naturally, now we have $\langle0|=[1\,\,0]$ and $\langle1|=[0\,\,\,1]$
and a NOT gate (also called Pauli $X$ gate) which transforms state
$|0\rangle$ to $|1\rangle$ and $|1\rangle$ to $|0\rangle$ may
be written as $X=|0\rangle\langle1|+|1\rangle\langle0|=\left[\begin{array}{cc}
0 & 1\\
1 & 0
\end{array}\right],$ where the first (second) term represents the transition of quantum
state $|1\rangle$ to $|0\rangle$ ($|0\rangle$ to $|1\rangle$).
It's analogous to the conventional NOT gate. Other single-qubit gates
often encountered are Hadamard gate, represented by $H=\frac{1}{\sqrt{2}}\left[\begin{array}{cc}
1 & 1\\
1 & -1
\end{array}\right]$ (this gate transforms $|0\rangle$ to $\frac{1}{\sqrt{2}}\left(|0\rangle+|1\rangle\right)$
and $|1\rangle$ to $\frac{1}{\sqrt{2}}\left(|0\rangle-|1\rangle\right)$),
phase gate $P\left(\phi\right)=\left[\begin{array}{cc}
1 & 0\\
0 & \exp\left(i\phi\right)
\end{array}\right],$ which introduces a relative phase. It can be visualized easily, if
we apply $P\left(\phi\right)$ on an arbitrary 1-qubit state $|\psi\rangle=\alpha|0\rangle+\beta|1\rangle=\left[\begin{array}{c}
\alpha\\
\beta
\end{array}\right]$, the output state will be $|\psi^{\prime}\rangle=P(\phi)|\psi\rangle=\left[\begin{array}{cc}
1 & 0\\
0 & \exp\left(i\phi\right)
\end{array}\right]\left[\begin{array}{c}
\alpha\\
\beta
\end{array}\right]=\left[\begin{array}{c}
\alpha\\
\beta\exp\left(i\phi\right)
\end{array}\right]=\alpha|0\rangle+\beta\exp(i\phi)|1\rangle.$ A special case of the phase gate is $Z$ gate $Z=P\left(\pi\right)=\left[\begin{array}{cc}
1 & 0\\
0 & -1
\end{array}\right].$ Other special cases are $S=P(\frac{\pi}{2})$ and $T=P(\frac{\pi}{4}),$
here it may be noted that along with a set of other gates\footnote{Specifically, gates from the family of Clifford group of gates and
$T$ gate.}, $S,\,S^{*},\,T,\,T^{*},X,\,Z$ and $H$ can be implemented directly
in the IBM quantum computers. A set of universal gates for quantum
computing requires all single qubit gates and at least one two qubit
gate. A widely applicable two qubit gate is a controlled-NOT (CNOT)
gate represented by ${\rm CNOT}=\left[\begin{array}{cccc}
1 & 0 & 0 & 0\\
0 & 1 & 0 & 0\\
0 & 0 & 0 & 1\\
0 & 0 & 1 & 0
\end{array}\right],$ which corresponds to an operation that maps $|x\rangle|y\rangle\rightarrow|x\rangle|x\oplus y\rangle.$
Thus, the input-output relation for this gate is $|00\rangle\rightarrow|00\rangle,\,|01\rangle\rightarrow|01\rangle,\,|10\rangle\rightarrow|11\rangle,\,|11\rangle\rightarrow|10\rangle$\footnote{In analogy to the single qubit states, these states can be written
in matrix form as $|00\rangle=\left[\begin{array}{c}
1\\
0\\
0\\
0
\end{array}\right],\,|01\rangle=\left[\begin{array}{c}
0\\
1\\
0\\
0
\end{array}\right],\,|10\rangle=\left[\begin{array}{c}
0\\
0\\
1\\
0
\end{array}\right],\,|11\rangle=\left[\begin{array}{c}
0\\
0\\
0\\
1
\end{array}\right].$ These matrices can be used to visualize that the given matrix for
${\rm CNOT}$ actually corresponds to the given input-output map. }. Clearly, it flips the second bit when first bit value is $1$ and
keeps it unchanged otherwise. Thus, the first qubit controls what
would happen on the second. This is why the first qubit is called
control qubit and the second one is called the target qubit, and as
a whole the gate is referred to as the controlled-NOT or CNOT gate.
This gate can also be implanted directly in IBM quantum computers,
but there are some restrictions on which positions of a circuit a
CNOT gate can be applied. Such restrictions arise from the architecture
of the quantum computer, and the same will be discussed in Section
\ref{sec:IBM-Quantum-Computer}. Here it would be apt to note that
a quantum circuit is composed of one or more such quantum gates arranged
sequentially to accomplish certain desired task. In what follows,
we will introduce some optical counterparts of the building blocks
discussed here before we start describing some experiments with IBM
quantum computer.

\subsubsection{Understanding the concepts of quantum mechanics using the concepts
of optics \label{subsec:Understanding-the-concepts}}

\begin{figure}
\centering{}\includegraphics[scale=0.5]{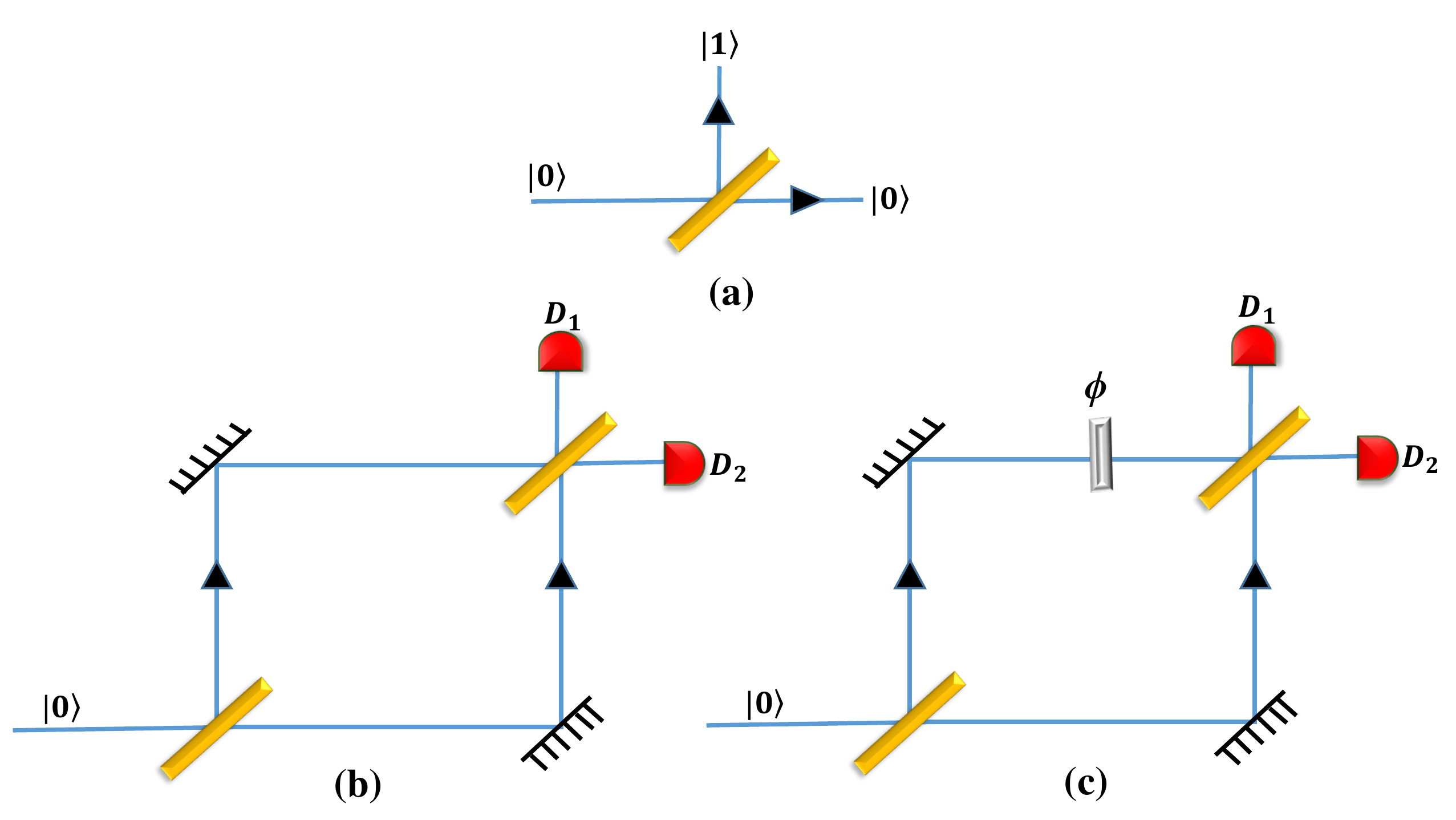}\caption{\label{fig:opt}(Color online) A single photon incident at the BS
either reflects with certain probability or transmits with the remaining,
which provides an optical illustration of qubit in (a). In this case,
the outputs travel in perpendicular directions and thus can be interfered
at another BS with the help of two mirrors (as shown in
(b)), known as MZI. In a standard Mach-Zehnder
interferometer with symmetric BSs only one of the detectors
clicks, while inserting a phase plate of thickness $t$ in one of
the arm (equivalent to applying a phase gate with angle $\phi=\frac{2\pi\left(n-1\right)t}{\lambda}$
with $n$ refractive index of the medium and $\lambda$ wavelength
of light used) the clicks on the other detector can be obtained. Here, the BS operation is equivalent to the Hadamard gate.}
\end{figure}

To elaborate on the idea of qubit, quantum gate and quantum circuit,
we will now use some basic optical elements. To begin with, let us
consider a symmetric BS, which reflects one half of the total number
of incoming photons while transmitting the other half.  To comprehend the idea of qubit, suppose a
single photon source emits a photon (represented by $|0\rangle$),
which falls upon a BS (at an incident angle of 45$^{{\rm o}}$)
to split into two output arms with equal probability amplitude as
$\frac{1}{\sqrt{2}}\left(|0\rangle+|1\rangle\right),$ where the photon
in the transmitted path is one level of the system denoted by $|0\rangle$
and photon in the other possible path, i.e., reflected path is the
other allowed level of our 2-level system which is represented by
$|1\rangle$ (as shown in Fig. \ref{fig:opt} (a)). So far we have
discussed a symmetric BS, i.e., a BS which reflects and transmits
photons with equal probability. For our discussion, such a symmetric BS can be considered as equivalent to Hadamard operation. Now, we may consider an asymmetric BS,
which transmits (reflects) photons with transmittance
(reflectance) $\alpha\,\left(\beta\right)$ such that
$\left|\alpha\right|^{2}+\left|\beta\right|^{2}=1,$ we can obtain
the output as an optical qubit $|\psi\rangle=\alpha|0\rangle+\beta|1\rangle.$ 

Coming back to our discussion on the symmetric BS, the
two outputs in Fig. \ref{fig:opt} (a) travel in the orthogonal directions.
Therefore, if we wish to interfere them we may need to use two mirrors
to direct these outputs as inputs of another BS (as shown
in Fig. \ref{fig:opt} (b)). At
the second BS the input wavepackets from two orthogonal
arms interfere constructively at one output and destructively at the
other one, which can be verified by putting one detector on each output
path and observing that only one of them (D$_{2}$) clicks. One can
easily show this using some simple matrix products. 

The state after the first BS is $H|0\rangle=\frac{1}{\sqrt{2}}\left[\begin{array}{cc}
1 & 1\\
1 & -1
\end{array}\right]\left[\begin{array}{c}
1\\
0
\end{array}\right]=\frac{1}{\sqrt{2}}\left[\begin{array}{c}
1\\
1
\end{array}\right]=|+\rangle.$ As mirrors are installed at both the arms, we can neglect its contribution
as a global phase and note that Hadamard is a Hermitian (i.e., self-inverse)
gate, thus we obtain the state after it passes through the second
BS as $H|+\rangle=\frac{1}{2}\left[\begin{array}{cc}
1 & 1\\
1 & -1
\end{array}\right]\left[\begin{array}{c}
1\\
1
\end{array}\right]=\left[\begin{array}{c}
1\\
0
\end{array}\right]=|0\rangle.$ Therefore, one of the detectors at the output of the second BS
would always click. It is straightforward to understand that such
arrangement of simple optical elements to show interference can be
referred to as an optical circuit.

Now, suppose we place a transparent plate of thickness $t$ and refractive
index $n$ in one of the output paths of the first BS (for
convenience choose the transmitted path as shown in Fig. \ref{fig:opt}
(c)). It would be equivalent to the application of the phase gate
$P(\phi)$ as this glass plate would introduce a relative phase shift
$\phi=\frac{2\pi\left(n-1\right)t}{\lambda}$, which depends on the
parameters such as thickness $t$, refractive index of the medium
$n$, and $\lambda$ wavelength of light used. We can obtain the output
of MZI in this case as $HP\left(\phi\right)H|0\rangle=\left[\begin{array}{c}
\cos\frac{\phi}{2}\\
-i\sin\frac{\phi}{2}
\end{array}\right].$ For $\phi=0$, we can obtain the results when phase plate was not
present. In what follows, we will perform the experiment to show the
same result. 

\section{IBM Quantum computer \label{sec:IBM-Quantum-Computer}}

As we mentioned in the above, qubits can be realized in various ways.
In case of the quantum computer placed in cloud by IBM, qubits are
realized using Josephson junction. Specifically, the superconducting
qubits used in the IBM architecture \cite{IBM-qubits-review} are known
as Transmon qubits. 

\begin{figure}
\centering{}\includegraphics[scale=0.5]{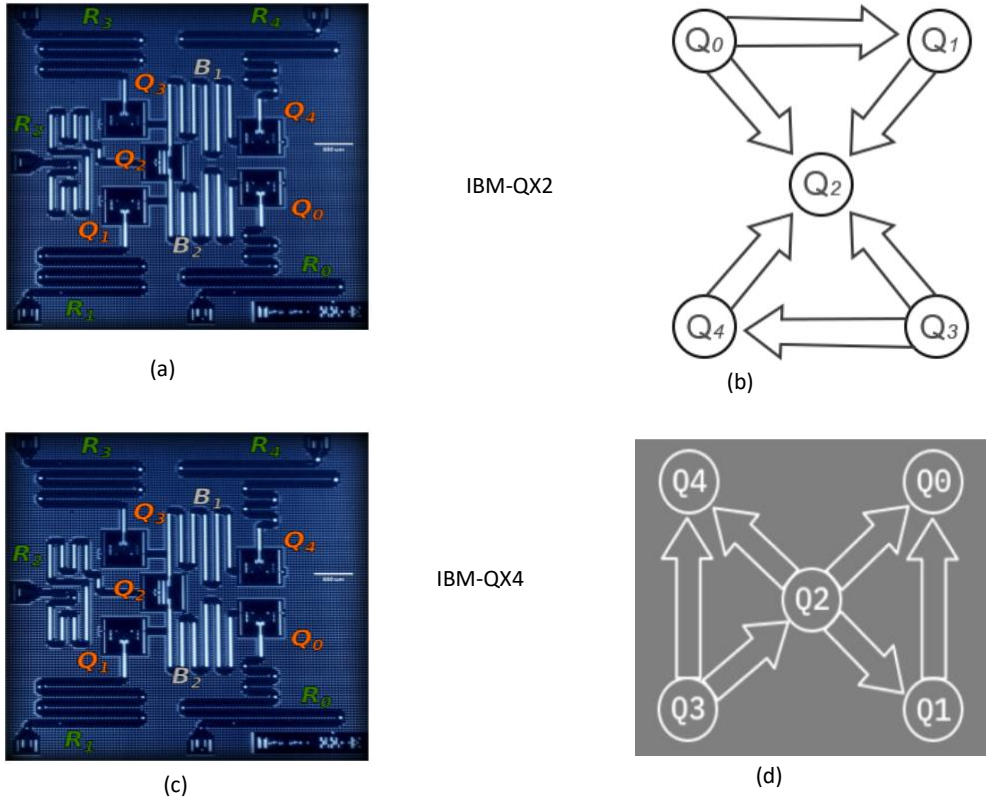}\caption{\label{fig:IBM-CNOT}(Color online) Two different architectures of
five qubit IBM quantum experience. The difference can be seen from
the allowed directions of CNOT. For instance, a CNOT is allowed with
control on Q0 (Q2) and target on Q2 (Q0) in IBM-QX2 (IBM-QX4), respectively. }
\end{figure}

There are 3 different architectures of computer that are available
and can be accessed freely through the internet. The one which was introduced
first was known as IBMQX2. Subsequently, two other architectures IBMQX4
and IBMQX5 have been introduced. IBMQX2 and IBMQX4 are 5 qubit quantum
computers, whereas IBMQX5 is a 16 qubit quantum computer\footnote{IBM is also providing a 20 qubit quantum computer QS1\textunderscore1 available to hubs, partners, and members of the IBM network. Therefore, we are not going to discuss QS1\textunderscore1 in this work.}.
In these computers, we are not allowed to implement any arbitrary
gate. We have to select gates from a library of gates which is comprised
of the gates from Clifford group and $T$ gate. Specifically, it contains
three Pauli\footnote{Three Pauli gates are NOT gate $X$, phase gate $Z=P\left(\pi\right)$, and $iY=X*Z.$}, identity, Hadamard, phase gates $S\,\left(S^\dagger\right)$ and $T\,\left(T^\dagger\right)$ as single qubit gates and ${\rm CNOT}$ as 2-qubit
gate. It also allows single qubit operations $U1=P\left(\phi\right)$, $U2=\frac{1}{\sqrt{2}}\left[\begin{array}{cc}
1&-\exp(i\phi)\\
\exp(i\Phi)&\exp(i\phi+i\Phi)
\end{array}\right]$, and $U3=\left[\begin{array}{cc}
\cos\frac{\theta}{2},-\exp(i\phi)\sin\frac{\theta}{2} \\ \exp(i\Phi)\sin\frac{\theta}{2},\exp(i\phi+i\Phi)\cos\frac{\theta}{2}\end{array}\right]$.  One can apply single qubit gates at any desired
point in the quantum circuits to be built using IBM quantum experience.
However, the application of CNOT gate (i.e., the positions in the quantum circuit, where CNOT gate can be applied) is restricted.
To be precise, in Fig. \ref{fig:IBM-CNOT}, we can see that there
are certain arrows and the position and direction of the arrows distinguish
IBMQX2 and IBMQX4. A particular arrow indicates that a CNOT gate can
be applied using the qubit shown at the tail of the arrow as the control
qubit and the qubit shown at the head of the arrow as the target qubit.
A CNOT is allowed with control on Q0 (Q2) and target on Q2 (Q0) in
IBMQX2 (IBMQX4), but the same is not allowed in IBMQX4 (IBMQX2). Actually, the interaction between the qubits are not the same for
all choices of two qubits. In fact, the interaction between the qubits
is stronger when a qubit having higher frequency is selected as the
control qubit, and qubit having lower frequency is chosen to be the
target. Thus, the frequencies of the qubits determines the directions
in which CNOT gates can be applied directly. In other words, these
frequencies lead to the arrows shown in Fig. \ref{fig:IBM-CNOT} and
thus the architecture of a particular implementation of IBM quantum
computer. This map (the architectures shown in Fig. \ref{fig:IBM-CNOT})
can be written in a compact form. For example, for the IBMQX2 the
connectivity map is given by coupling\_map = \{0: {[}1, 2{]}, 1: {[}2{]},
3: {[}2, 4{]}, 4: {[}2{]}\}, where a: {[}b{]} means a CNOT with qubit
a as control qubit and b as target qubit can be implemented. See that
this map describes the architectures shown in Fig. \ref{fig:IBM-CNOT}
(b). Similarly, the map for IBMQX4 is coupling\_map = \{1: {[}0{]},
2: {[}0, 1, 4{]}, 3: {[}2, 4{]}\}. From these coupling maps also (or
equivalently from the architectures shown in Fig. \ref{fig:IBM-CNOT}
(b) and (d)), one can easily recognize that application of CNOT from
Q0 to Q2 (Q2 to Q0) is allowed in IBMQX2 (IBMQX4), but is not allowed
in IBMQX4 (IBMQX2).

Being superconductivity-based quantum computers, all of these IBMQX{*}
work in very low temperature. Specifically, operating temperature
for IBMQX2 and IBMQX4 are 0.0178 K and 0.021K, respectively.

\subsection{How to use IBM quantum computer}

\begin{figure}
\centering{}\includegraphics[scale=0.5]{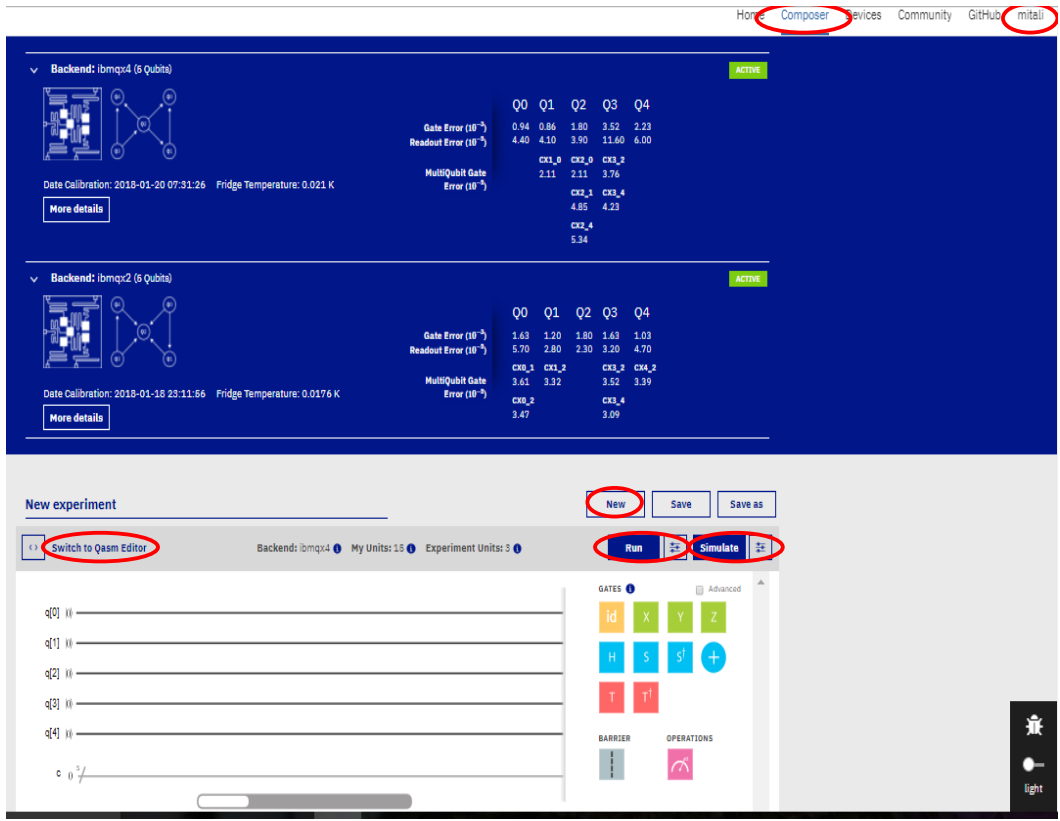}\caption{\label{fig:Comp}(Color online) The window that we see after clicking
the ``Composer'' tab, where different quantum circuits can be built
using drag and drop.}
\end{figure}

To start using the IBM quantum computer, you have to first register
yourself and thus create a login and password. To do the same please
follow the following steps:
\begin{enumerate}
\item Open the website of IBM Q Experience, i.e., open https://quantumexperience.ng.bluemix.net/qx/experience
\cite{IBM}.
\item Depending on the web-browser used by you either a ``Sign in'' popup
window will open automatically, or you have click on the ``Sign in''
button on the top right corner which will open a new tab. Subsequently,
first time users have to click on ``Sign up'' button. 
\item An account will be created after filling all the required fields. 
\end{enumerate}
In this process, your email-id will be your login id and the password
is of your choice. Alternatively, one can also login using one of
his/her ids of Linkedin, Github, Twitter, etc. After performing registration
go to the website \cite{IBM} and sign in with your credentials. At
the top-right corner you will see your name as entered during the
registration process. Once you sign in, follow the follwing steps.
\begin{enumerate}
\item Click on the ``Composer'' tab and select a topology that you wish
to use for your experiment.\\
\textbf{Note:} After clicking on the ``Composer'' tab, you will
be prompted to choose among three available topologies of the IBM
quantum computers, namely IBMQX4, IBMQX2, and Custom topology. To
run your results on a real quantum computer choose either IBMQX4 or
IBMQX2, while Custom topology is useful for running simulation of
your circuits. Infact, clicking on the Composer tab (marked at the
top in Fig. \ref{fig:Comp}), you will see a window as shown in Fig.
\ref{fig:Comp}. You can see that the user name is appearing in the
top right. It's ``mitali'' in our example. In your case, it will
be replaced by your name. \\
\item Select appropriate gates and create your quantum circuit.\\
\textbf{Note:} In the lower side of Fig. \ref{fig:Comp} (i.e., in
the lower portion of the window that you have opened in the previous
step), you can see five horizontal lines (indexed by ${\rm q[0],\,q[1],\,q[2],\,q[3]}$and
${\rm q[4],}$respectively) representing five qubit lines, on which
different unitary operations (quantum gates) allowed in IBM quantum
computer can be dropped/dragged after selecting them from the set
of gates shown in the right hand side of the qubit lines (see right
side of the window/figure). The last line (i.e., the line at the bottom
which is indexed by ${\rm c_{0}}$) corresponds to the classical registrar
which would store the classical values of the measurement outcomes.
In the figure as well as in the window you that you have opened in
IBM quantum experience, you can see that all the 5 qubits are initially
prepared in the state $|0\rangle.$ Thus, the initial state of an
IBM quantum computer is always $|00000\rangle$. Further, you can
see that the choice of topology you have made in the previous step
is clearly mentioned over the qubit lines as Backend, where units
assigned to a user for using the quantum computer are also mentioned.
In its left, one can see text ``Switch to QASM Editor'', which allows
one to design the quantum circuit by writing a program in QASM. We
will briefly discuss it in the forthcoming section. 
\item Either run or simulate the circuit that you have designed in the previous
step. To do so, click on the corresponding tabs (i.e., click either
on ``Run'' tab or on ``Simulate'' tab). \\
\textbf{Note:} To run or simulate the circuit and to see its outcome
you have to perform measurement on the appropriate qubits. Once you
have designed a circuit and performed measurements on suitable qubits,
you will receive your measurement outcomes in the computational basis, i.e., $\left\{|0\rangle,|1\rangle\right\}$.
As mentioned above, you can either choose to simulate the output of
the circuit or run it on the quantum computer. Next to ``Run'' and
``Simulate'' tabs, there are tabs which can be used to change the
number of shots you wish to run to obtain the probability distribution
of the output. The higher the number of shots the more units are required
to run the IBM quantum computer. However, with increase in the number
of shots we obtain better results. This point will be established
in the next section with the help of an example.
\item In the previous step, the experiment will be performed and you will
obtain the result either immediately in a new window (for simulation,
we will always obtain it immediately) or the job will be placed in
queue and you will receive an email from IBM when the job is executed.
Subsequently, you can login again and see the result of your experiment.
\end{enumerate}
In case of difficulty, one can also refer the IBM tutorials available for beginners \cite{IBM-help}. 

\section{Performing simple experiments with IBM quantum computer to clarify
the concept of quantum mechanics \label{sec:Performing-simple-experiments}}

\subsection{Experiment 1: Is quantum mechanics probabilistic?}

\begin{figure}
\centering{}\includegraphics[scale=0.5]{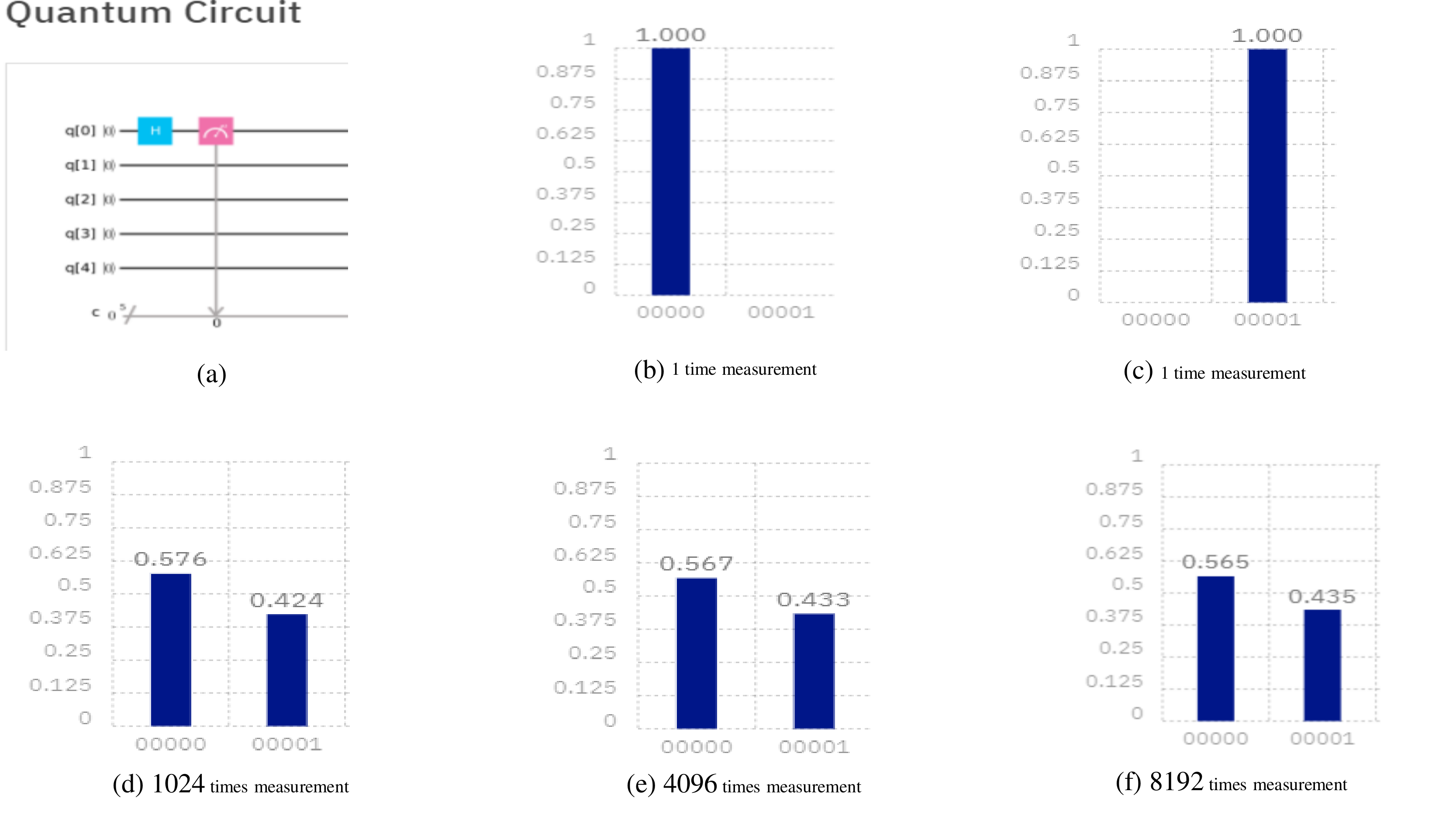}\caption{\label{fig:H-meas}(Color online) Randomness intrinsic in quantum
mechanics is shown with the help of measuring an equal superposition
state in the computational basis (as shown in (a)) in (b) and (c)
for single run. However, probabilistic interpretation that we use
requires higher number of runs. With increase in the values of runs
we can observe equal probabilities for both measurement outcomes.}
\end{figure}

We have already introduced the idea of quantum computation and tools
that we are going to use to perform our experiments. To begin with,
we will try to develop a feeling about the probabilistic nature of
quantum mechanics. For this purpose, we may aim to prepare a quantum
state $|\psi\rangle=\frac{1}{\sqrt{2}}\left(|0\rangle+|1\rangle\right)$
which is in equal superposition of bit values 0 and 1. In the above
we have seen that this state can be obtained as the output of a beam
splitter with single photon input (cf. Fig. \ref{fig:opt} (a)). In
case of IBM quantum computers, the equivalent system can be designed
by placing a Hadamard gate in any qubit line. As the Hadamard maps
$|0\rangle$ to $\frac{1}{\sqrt{2}}\left(|0\rangle+|1\rangle\right),$
and as the default initial (input) state for each qubit line is $|0\rangle$ in
the IBM quantum computers, the output of the Hadamard gate will be
in the state $|\psi\rangle=\frac{1}{\sqrt{2}}\left(|0\rangle+|1\rangle\right)$.
Now, we perform measurement, choosing number of shots to be 1 (which
implies that the measurement is performed only once, i.e., the experiment
is not repeated). How many times, you wish to perform the same experiment
(repeat the experiment) can be chosen by clicking on the button in
the right side of Run button and subsequently clicking on Edit parameters).
The single shot measurement would correspond to the situation where
a single quantum state $\frac{1}{\sqrt{2}}\left(|0\rangle+|1\rangle\right)$
is measured in the computational basis (circuit is shown in Fig. \ref{fig:H-meas}
(a)). Interestingly, in a single run, one can either obtain measurement
outcome $|0\rangle$ or $|1\rangle$ as we have shown in Fig. \ref{fig:H-meas}
(b) and Fig. \ref{fig:H-meas} (c) which are obtained as outcomes
in different runs. Here it's important to note that the measurement outcome of the IBM quantum computers are to be read from the right to left. The right most bit value corresponds to the outcome of measurement on q[0], next one in the left corresponds to the measurement on q[1], and so on. For example, the outcome of the measurement performed in the circuit shown in Fig. \ref{fig:H-meas} (a) is shown in Fig. \ref{fig:H-meas} (c) as a single bar at 00001. The last digit shows that the measuring q[0] we have obtained 1. As neither any operation nor any measurement is performed on the other qubits output state for them is shown as 0.
If we keep on repeating this single shot experiments,
we will obtain a random sequence of 0 and 1. This is what happens
in the quantum random number generators. Repeated execution of such
single shot experiment would convince you that on measurement the
wave function (or the state $|\psi\rangle$) randomly collapses to
one of the allowed states (in this case either in $|0\rangle$ or
in $|1\rangle$). Thus, we have demonstrated the phenomenon of wave
function collapse on measurement which is a distinguishing feature
of quantum mechanics.

After showing collapse of wavefunction on measurement, we may perform
the experiment again with higher numbers of shots, and in Fig. \ref{fig:H-meas}
(d)-(f), one can observe that for higher values of shots we have obtained
probability distribution of the measurement outcomes. We have already
discussed how to select number of shots. If we choose 8192 shots,
the experiment will be repeated 8192 times, and in the result probability
of obtaining 0 and 1 will be shown. See with the increase in the number
of shots i.e., number of times an experiment is repeated, we approach
closer to verify the statement that on measurement (in computational
basis) a quantum state $|\psi\rangle=\alpha|0\rangle+\beta|1\rangle$
collapses to state $|0\rangle$ and $|1\rangle$ with probabilities
$|\alpha|^{2}$ and $|\beta|^{2},$ respectively. In the particular
case for which experiment is performed here, we have $\alpha=\beta=\frac{1}{\sqrt{2}}$,
and consequently corresponding probabilities are $\frac{1}{2}$ in
the ideal situation. In reality, we observe that unless the experiment
is repeated a large number of times, the probabilities would not approach
$\frac{1}{2}$ (that's natural in any statistical event). In other
words, this shows the statistical nature of quantum mechanics. Further,
even for 8192 shots, the probabilities are not exactly $\frac{1}{2}$.
This is so partially because of two reasons, (i) even 8192 is not
a statistically large number, and (ii) there may be some errors which
can be attributed to noise in the quantum system or gate errors.

\subsubsection{Application of Experiment 1: Quantum random number generator}

The experiment discussed here not only establishes the probabilistic
nature of quantum mechanics, but also forms the basis for generating
a string of true random numbers. Note that there does not exist a
true random number generator in the domain of classical physics, while
due to inherent randomness of quantum mechanics a quantum random number
generator can be built. The fact that classical random numbers are
not truly random, can be easily visualized through a lucid example.
Consider the outputs of repeated tossing of a coin. Usually we would expect
the outputs to be random. However, if we know the air drag, weight
of the coin,  force applied at the time of throwing the coin, height at the time of throwing,
gravitational acceleration, etc., we can in principle solve the equation
of motion and predict the output. So the output is not random. Here,
randomness actually arises due to our ignorance. In contrast, quantum
mechanics is intrinsically probabilistic. 

The working of an easily available quantum random number generator
involves simple optical elements- a symmetric BS and two detectors.
Specifically, performing measurement in the two output ports of the
BS in Fig. \ref{fig:opt} (a) one can obtain a string of 0s and 1s.
Corresponding experiment performed on IBM quantum experience with
shot value 1 as shown in Fig. \ref{fig:H-meas} (a) gives random outputs
0 and 1 as in Fig. \ref{fig:H-meas} (b) and (c), respectively. A
repetitive preparation of this initial state and its measurement in
the computational basis can be used to obtain a string of true random
numbers. Subsequently, an interested reader can perform various randomness
tests to ensure that the generated numbers are random. See
\cite{rand} for the tools for various types randomness tests recommended
by NIST. Here, it may be noted that random number generators are used
in some ATMs, Casinos, in Weather predictions, stock-market predictions,
etc., and in many other places including state lottery boards. In
brief, the simple experiment described above not only illustrates
a distinct character of quantum mechanics, it also establishes quantum
supremacy in context of a particular application of quantum mechanics
that has relevance in our day to day life.

\subsection{Experiment 2: Mach-Zehnder interferometer }

\begin{figure}
\centering{}\includegraphics[scale=0.5]{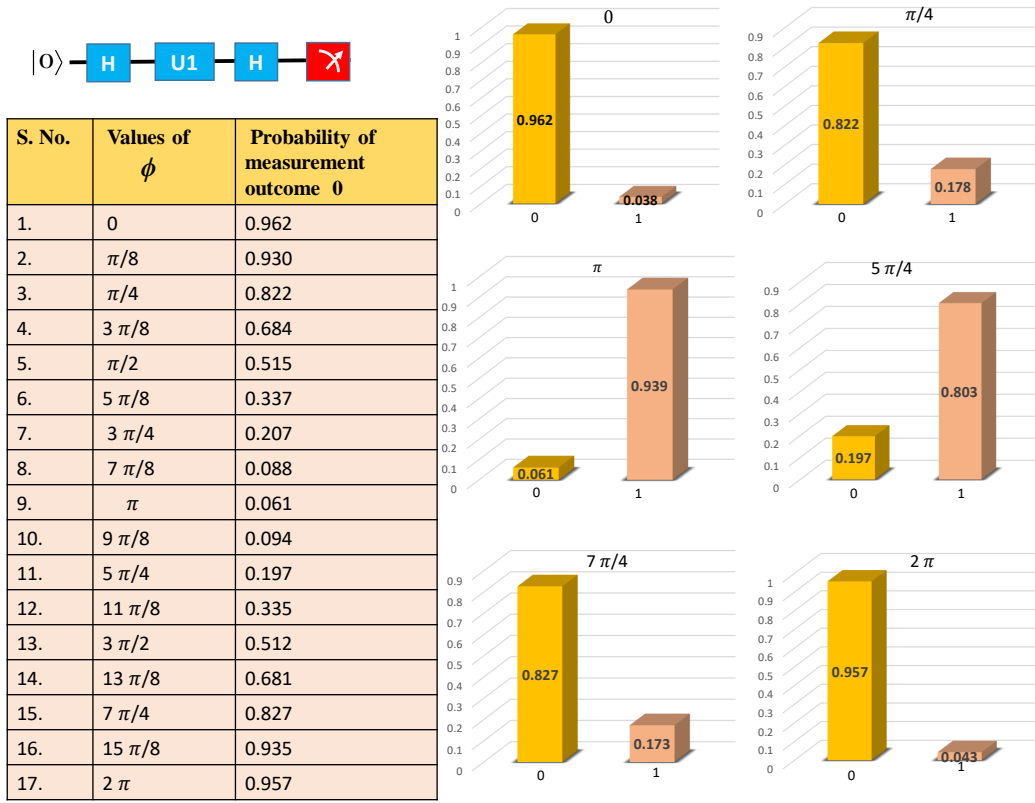}\caption{\label{fig:MZI}(Color online) MZI realized in IBM quantum experience.
Probabilities of obtaining measurement outcome 0 for different values
of the phase angle $\phi$ are summarized in table. Few particular
cases (for specific values of $\phi$) as output of IBM quantum
experience are also shown.}
\end{figure}

\begin{figure}
\centering{}\includegraphics[scale=0.35]{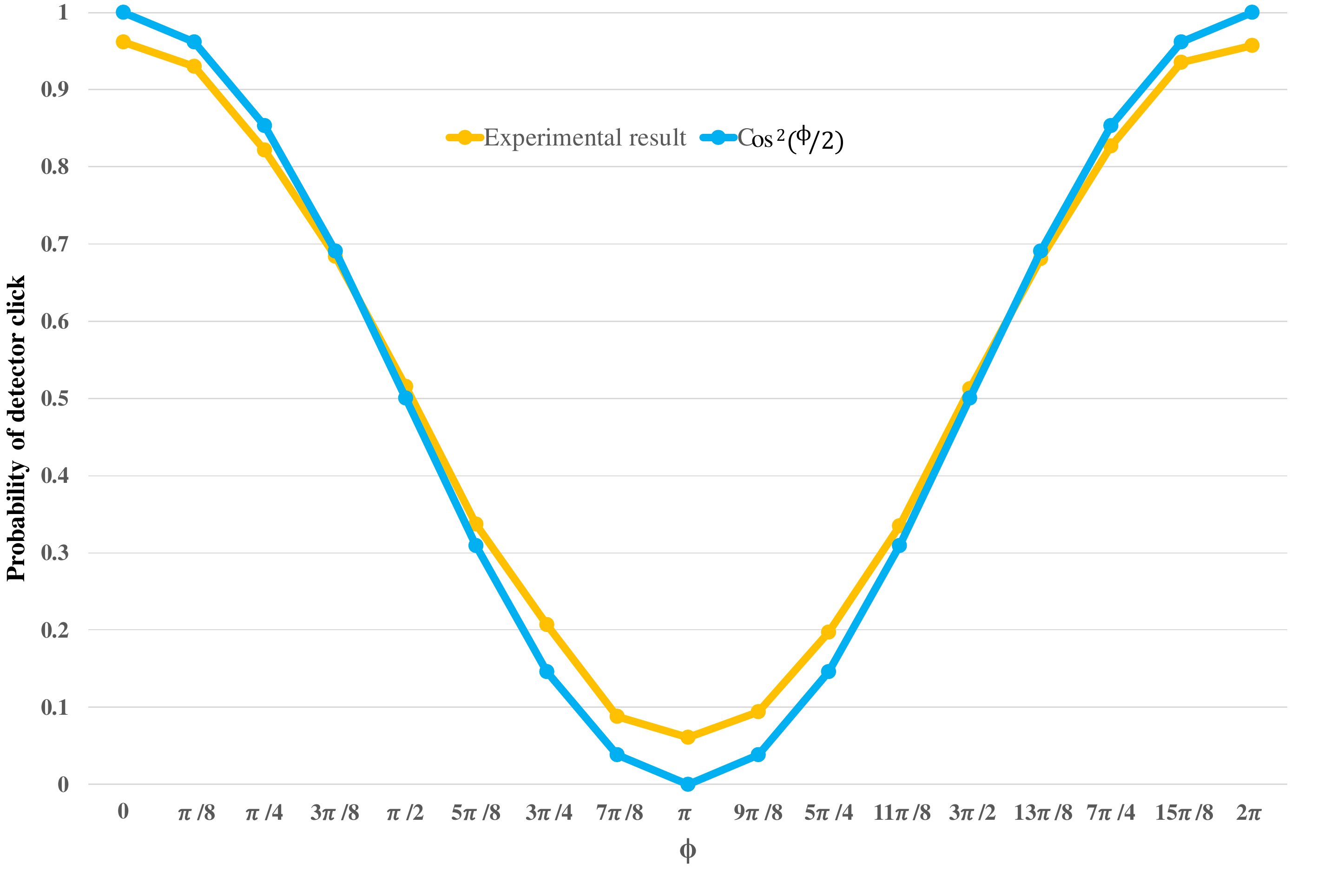}\caption{\label{fig:Exp-result}(Color online) The experimental output of MZI (realized using IBM quantum computer) is shown to
fit nicely with the corresponding theoretical results.}
\end{figure}

We have discussed MZI while discussing optical
circuits. We have shown that due to insertion of a phase plate (with phase angle
$\phi$), the output bit values 0 and 1 are obtained with probabilities
$\cos^{2}\frac{\phi}{2}$ and $\sin^{2}\frac{\phi}{2}$, respectively.
For our experiments Hadamard gate works like a BS, and consequently
applications of 2 consecutive Hadmard gates will be equivalent to
a Machzehnder interfeorometer as it would imply the use of output
of first BS as the input of second BS. The role of mirrors in the original
MZI is just to redirect the output of the
first BS to the input ports of the second BS, so in IBM quantum computer,
we don't need any component (gate) analogous to mirror. Now, a Mach-Zehnder
interferometer with a phase plate in one of the path (as shown in 
Fig. \ref{fig:opt}) should be equivalent to a circuit in which a phase gate
is inserted between two Hadamards. Such a circuit designed for the
implementation in IBM quantum computer is shown in the top-left corner
in Fig. \ref{fig:MZI}. We performed the experiment using this circuit
in IBM quantum computer with different values of $\phi$ and prepared
a table of values of probability of measurement outcome 0. Some of
the obtained results are also shown Fig. \ref{fig:MZI} for illustration.
We obtained the variation of probability of detector click (correspond
to bit value 0) and have shown it to fit with the plot of $\cos^{2}\frac{\phi}{2}$
(cf. Fig. \ref{fig:Exp-result}). Thus, our experimental results match
exactly with the theoretically calculated value. 

Here it is important to note that the first experiment performed here
(which led to random number generator) does not establish the the
fact that a quantum state simultaneously existed in state $|0\rangle$
and $|1\rangle$sates. Consider a model which tells that a quantum
particle after interacting with the BS randomly goes to the reflected
path in 50\% cases and in the transmitted cases in the rest of the
cases. In that case also the detectors placed after the BS
(see Fig. \ref{fig:opt}) would have clicked randomly. So the
previous experiment cannot distinguish between this theory and the
theory that tells that a quantum state simultaneously remains in $|0\rangle$
and $|1\rangle.$ However, as soon as we add the next BS in the MZI,
this theory would imply that independent of the fact whether the quantum
particles come from the reflected path or the transmitted path, half
of them will go to one detector after the second BS and rest will
go to the second detector. However, if the interpretation of quantum
mechanics which states that the quantum particle simultaneously stays
in both the paths is correct then constructive interference will happen
in one of the detectors (which will always click) and destructive
interference will happen on the other detector (which will never click).
Applying two consecutive Hadamards in IBM, we can demonstrate that
the quantum states really stay in the superposition state and the
crude model described in this paragraph is wrong.

\subsubsection{Application of Experiment 2: Interaction free measurement and quantum
cryptography}

In this experiment, we have seen that MZI can be implemented using
IBM quantum computer. It's interesting for various reasons. Specifically,
MZI and its variants have been used to realize numerous quantum communication
and computation schemes. For instance, Goldenberg-Vaidman \cite{GV}
and Guo-Shi \cite{Guo-Shi} protocols of quantum key distribution
essentially use MZI (see Chapter 8 of Ref. \cite{book} for details).
Further, counterfactual quantum communication proposed in the recent
past \cite{salih} and experimentally performed recently \cite{exp}
employ chained MZI. 

\subsection{Experiment 3: Prepare and visualize an entangled state}

\begin{figure}
\centering{}\includegraphics[scale=0.3]{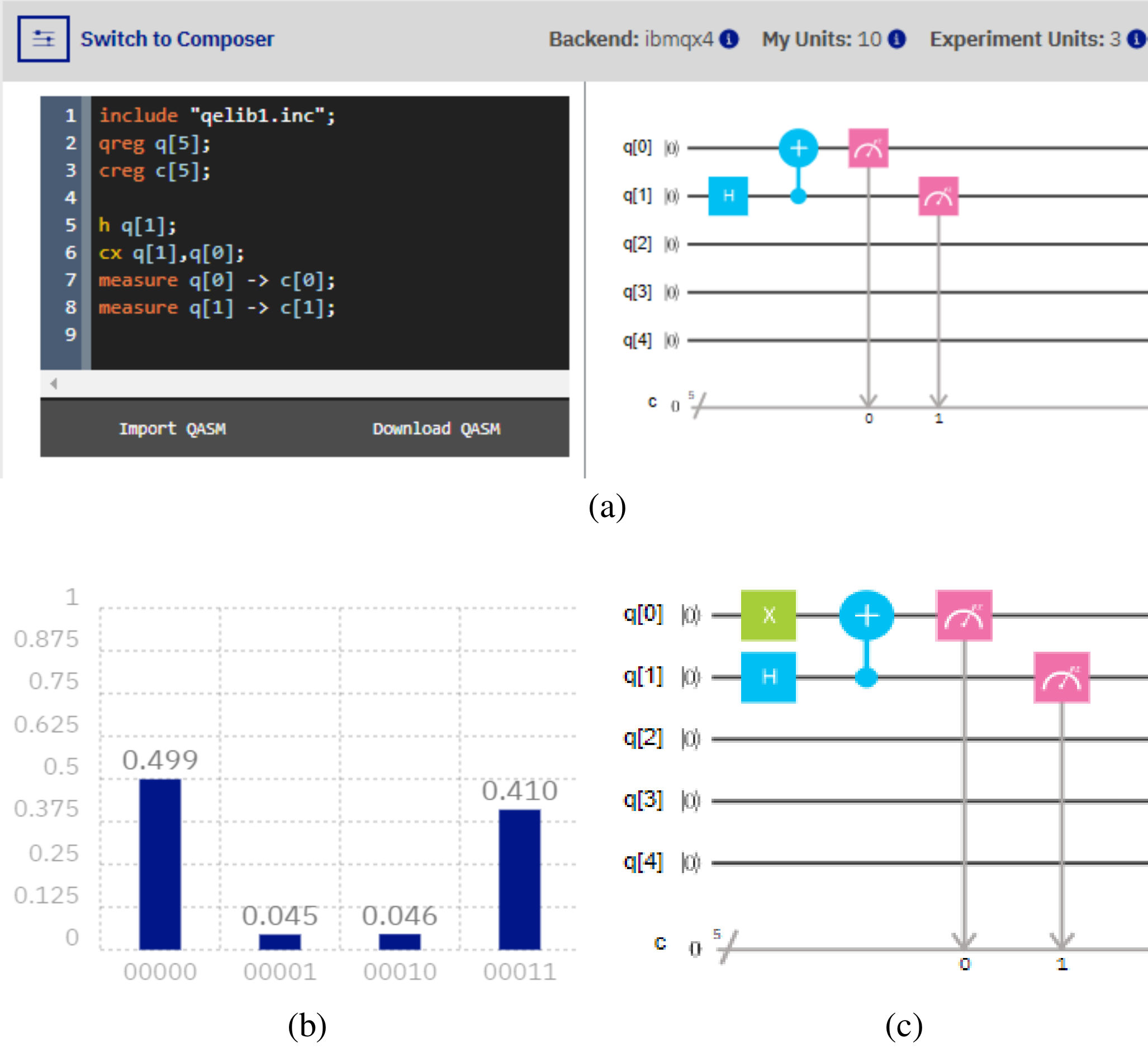}\caption{\label{fig:Exp-qasm}(Color online) In addition to drag and drop approach,
one can also use QASM to write the circuits. Here, as an example,
we have shown generation of entanglement using a Hadamard and a CNOT
gate. }
\end{figure}

{Suppose that there are two quantum systems indexed
by the subscripts $A$ and $B$ and they form a composite system indexed
by subscript $AB$. Thus, the systems indexed by the subscripts $A$
and $B$ can be considered as the subsystems of a bigger (composite)
system indexed by subscript $AB$. Further, consider that the quantum
state of the first subsystem is $|\psi\rangle_{A}$ and that of the
second subsystem is $|\psi\rangle_{B}$. Now, if we can write $|\psi\rangle_{AB}=|\psi\rangle_{A}|\psi\rangle_{B}$
then the composite state is called separable and otherwise it's called
entangled or inseparable. For a better understanding of entanglement
one has to read about tensor product (see Chapter 3 of \cite{AP}). However,
it's possible to develop a feeling of entangled state through some
simple examples. To provide a lucid idea, let us think that the subsystems
$A$ and $B$ are single qubit systems. Now if we have $|\psi\rangle_{AB}=\frac{1}{\sqrt{2}}\left(|00\rangle+|10\rangle\right)_{AB}$,
we can easily see that the second qubit is always in the state $|0\rangle$
and $|\psi\rangle_{AB}$ can be separated (factorized in a lucid sense)
in the form $|\psi\rangle_{AB}=|\psi\rangle_{A}|\psi\rangle_{B}$
as follows $|\psi\rangle_{AB}=\frac{1}{\sqrt{2}}\left(|00\rangle+|10\rangle\right)_{AB}=\frac{1}{\sqrt{2}}\left(|0\rangle+|1\rangle\right)_{A}|\psi\rangle_{B}$.
Thus, the state $|\psi\rangle_{AB}=\frac{1}{\sqrt{2}}\left(|00\rangle+|10\rangle\right)_{AB}$
is separable (as you can separate the states of the first and second
qubits). Now, you can see that following sates are not separable in
the above sense: $|\psi^{+}\rangle_{AB}=\frac{1}{\sqrt{2}}\left(|00\rangle+|11\rangle\right)_{AB}$
and $|\phi^{+}\rangle_{AB}=\frac{1}{\sqrt{2}}\left(|01\rangle+|10\rangle\right)_{AB}$.
Being inseparable, these states are called entangled. These states
have no classical analogue and the collapse on measurement described
and verified earlier leads to very interesting consequences for these
states. To begin with you can see that if you measure the first qubit
in the state $|\psi^{+}\rangle_{AB}=\frac{1}{\sqrt{2}}\left(|00\rangle+|11\rangle\right)_{AB}$
and obtain $|0\rangle_{A}$ then the other qubit must collapse to
$|0\rangle_{B}$. Similarly, if your measurement on the first qubit
yields $|1\rangle_{A}$ then the state of the second qubit must become
$|1\rangle_{B}$ . Thus, there is a kind of quantum correlation. Let
us now show you how to produce $|\psi^{+}\rangle$ and $|\phi^{+}\rangle$in
IBM quantum computer and what kind of outcome is obtained in the measurement.
However, for the completeness of the article, instead of using the
drag and draw approach, here we will follow} another method for preparing
the quantum circuits. To be precise, as mentioned above, after opening composure we have clicked on the "Switch to QASM Editor" tab and that has led us to a black window, where we have written a simple QASM code \footnote{It's easy to understand the code. First 3 lines are common in all the programs and appear automatically in the QASM editor, commands shown in line 2 (3) creates the quantum (classical) register. Line number 7 and 8 correspond to two measurements that are shown in the right side of the right panel of Fig. \ref{fig:MZI} (a). Line 5, commands to apply a hadamard gate (written as h, similarly $\rm{NOT}$ gate can be written as x) on the second qubit from the top, which is indexed as q[1]. Similarly, Line 6 of the program commands to apply a $\rm{CNOT}$ gate written as cx in a way that the second qubit from the top (i.e., q[1]) works as control qubit and the topmost qubit (i.e., q[0]) works as the target qubit. Now, in the vacant line 4 of the program, if we write x q[0] then we would obtain the circuit shown in Fig. \ref{fig:Exp-qasm} (c).} (see left panel of Fig. \ref{fig:Exp-qasm} (a) to generate the quantum circuit shown in Fig. \ref{fig:Exp-qasm} (b) which would produce a
two-qubit entangled state $|\psi^{+}\rangle$). On measurement, the output of this circuit is expected to be 00 or 11, so we can expect two bars of equal or almost equal height one at 00000 and one at 00011. However, by performing the real experiment, we found the output shown in Fig. \ref{fig:Exp-qasm} (b), where we can see a small but finite probability of obtaining 00001 and 00010, too. These, two small bars appears as a manifestation of gate errors and channel noise. Once, you see that even for a small circuit with only 2 gates, noise can affect the output to some extent, you can easily recognize what restricts us from building large (scalable) quantum computers. Finally, in Fig. \ref{fig:Exp-qasm} (c), we show a circuit that would produce the entangled state $|\phi^{+}\rangle$. 

We would like to
suggest the young readers to replace Hadamard gate with U3 gate in the circuits shown in Fig. \ref{fig:Exp-qasm} (a) and (c), and to obtain the corresponding results in the tabular form as was
shown in Fig. \ref{fig:MZI}. This exercise, would help them to understand the idea of maximally and
non-maximally entangled states.

\subsubsection{Applications of Experiment 3: Quantum key distribution, entanglement swapping and quantum
repeaters}

As mentioned previously, that the measurement outcomes of the entangled state $\frac{1}{\sqrt{2}} \left(|00\rangle+|11\rangle\right)$ in the computational basis are symmetric, two distant parties sharing multiple copies of this state can form a symmetric string of random numbers to be used as unconditionally secure quantum key. Starting with two copies of the entangled state and measuring both the first qubits in the Bell basis\footnote{Analogous to the computational basis, a two-qubit basis is defined as $\left\{|\psi^{\pm}\rangle=\frac{1}{\sqrt{2}} \left(|00\rangle+|11\rangle\right),|\phi^{\pm}\rangle=\frac{1}{\sqrt{2}} \left(|01\rangle+|10\rangle\right)\right\}$.}, the last two qubits get entangled. This is termed as entanglement swapping. The idea of entanglement swapping is useful in long distance quantum communication where entanglement between two distant parties can be shared with the help of measurements in Bell basis as quantum repeaters.

Here we have mentioned only a few simple applications of entanglement. 
Interested readers may read about quantum teleportation, dense coding,
remote state preparation, etc., which are more convincing applications
of entangled states. To be precise, none of these quantum phenomena
can be obtained without the use of quantum entanglement.

\section{Conclusion \label{sec:Conclusion}}

In the above, we have seen that the nonclassical features of quantum
mechanics can be tested through some simple experiments in the class
room, and such experimental realizations and their modifications have
direct applications in performing tasks having socio-economic relevance.
The experiments and applications mentioned here are only the representative
cases. Many such experiments can be designed and analyzed.  Such experiments
can also be used to teach advanced topics of quantum mechanics through
the experiments done using IBM quantum computer. Experimental studies also require one to compute fidelity of the quantum states, which can be reconstructed by quantum state tomography (see \cite{Fid} for detail). We conclude the article
with a hope that the interested teachers and students will try to
design new experiments with the help of this article and the texts
mentioned in the Further reading section, will provide the backbone
for such new designs. 

\section*{Further reading}
\begin{enumerate}
\item Optical quantum information and quantum communication, A. Pathak and
A. Banerjee, SPIE Spotlight Series, SPIE Press (2016) ISBN: 9781510602212
\item Light and its Many Wonders, A. Ghatak, A. Pathak and V. P. Sharma
(Eds.), Viva Books, New Delhi, India (2015) ISBN 978-81-309-3428-0
\item Beck, Mark. Quantum mechanics: theory and experiment. Oxford University
Press, 2012.
\end{enumerate}
\textbf{Acknowledgment: AP thanks Defense Research and Development Organization (DRDO), India for the support provided through the project number ERIP/ER/ 1403163 /M/01/ 1603. He also thanks K. Thapliyal and M. Sisodia for their technical feedback and help.}

\end{document}